# Single charge and exciton dynamics probed by molecular-scale-induced electroluminescence

A. Rosławska[1]*, P. Merino[1,+], C. Große[1,#], C. C. Leon[1],  O. Gunnarsson[1], M. Etzkorn[1,‡], K. Kuhnke[1], K. Kern[1,2]

[1] Max Planck Institute for Solid State Research, Heisenbergstraße 1, 70569, Stuttgart, Germany.

[2] Institut de Physique, École Polytechnique Fédérale de Lausanne, 1015 Lausanne, Switzerland.

+ present address: Instituto de Ciencia de Materiales de Madrid, CSIC, c/Sor Juana Inés de la Cruz 3, E28049, Madrid, Spain.

# present address: NanoPhotonics Centre, Cavendish Laboratory, University of Cambridge, Cambridge CB3 0HE, UK.

‡ present address: Institut für Angewandte Physik, TU Braunschweig, Mendelssohnstraße 2, 38106 Braunschweig, Germany

* corresponding author: a.roslawska@fkf.mpg.de

## ABSTRACT

**Excitons and their constituent charge carriers play the central role in electroluminescence mechanisms determining the ultimate performance of organic optoelectronic devices. The**

**involved processes and their dynamics are often studied with time-resolved techniques limited by spatial averaging that obscures the properties of individual electron-hole pairs. Here we overcome this limit and characterize single charge and exciton dynamics at the nanoscale by using time-resolved scanning tunnelling microscopy-induced luminescence (TR-STML) stimulated with nanosecond voltage pulses. We use isolated defects in $C_{60}$ thin films as a model system into which we inject single charges and investigate the formation dynamics of a single exciton. Tuneable hole and electron injection rates are obtained from a kinetic model that reproduces the measured electroluminescent transients. These findings demonstrate that TR-STML can track dynamics at the quantum limit of single charge injection and can be extended to other systems and materials important for nanophotonic devices.**

Organic light-emitting diodes (OLEDs) outperform inorganic-based LEDs in many ways such as flexibility, efficiency, and thickness. They are essential components of contemporary digital displays and have remarkable potential for illumination purposes[1-4]. The key to OLED technology is light emission due to the radiative recombination of electron-hole pairs (excitons) in organic thin films. This process is driven by applying a voltage across the macroscopic or mesoscopic film. The associated electroluminescence is, however, governed by the nanoscale dynamics of both, the excitons, and the charge carriers injected from the electrodes that sandwich the thin film. These processes can be followed and characterized with established techniques, which include pump-probe spectroscopy[5,6], time-of-flight methods[7] and time-resolved electroluminescence[8]. However, these techniques are mainly sensitive to bulk transport characteristics and rely on spatially-averaged measurements performed at high charge and exciton densities. Analogous measurements on organic films with nanometre-precise charge injection in the single charge and exciton regime are still awaited[5]. Understanding light emission dynamics at the quantum limit will clarify the factors that lead to higher electroluminescence conversion efficiency, in particular, the role of structural defects, organic-metal and organic-organic interfaces, and the influence of quantum coherence effects[5,9,10].

The goal of investigating charge and exciton dynamics at the molecular scale and in the limit of individual quanta calls for experimental advances that combine the manipulation of single electron-hole pairs on the length scale of 1 nm with the temporal resolution comparable to their characteristic decay times, which are on the order of 1 ns. Imaging molecular orbitals with sub-nm resolution[11-13] is routine with scanning tunnelling microscopy (STM). By extending this method to STM-induced luminescence (STML)[14,15], light emission from single exciton recombination events at isolated molecules[12,16-21] or single defects[22-24] can be recorded simultaneously. Moreover, the time resolution of STM measurements, conventionally limited by the millisecond response time of picoampere current amplifiers, can be circumvented by applying either short voltage[25-31] or laser pulses[32-34], with the latter reaching down to the femtosecond regime[35-39]. While laser-related techniques can achieve better time resolution, using fast voltage pulses as we do in this work, is preferable because it mimics the excitation mode of commercial OLEDs.

Contemporary OLED performance metrics are typically obtained by recording the electroluminescence response to an applied short voltage pulse[8]. We can now extend this technique to the molecular level by combining fast voltage pulses with STML and probe in the time domain the dynamics of injected charges that lead to exciton formation and their subsequent decay. This ability to track separately-created excitons arising from voltage pulsing is a fundamental strength of time-resolved STML (TR-STML). Demanded by our focus on the time domain, the electronic current in this work is often expressed in units of time per charge, which is equal to the average time interval between charge tunnelling events.

**Operating principle**

The system we study is a nanoscale model resembling a first-generation OLED comprising two metallic electrodes (tip and substrate) and a semiconducting $C_{60}$ thin film. While the optical transition between ground state and the lowest electronic excited state of $C_{60}$ is symmetry forbidden[40] structural defects (see Supplementary Fig. S1) known as X-traps[41] relax the selection rules[42] and permit the films to exhibit localized photo- and electroluminescence[43]. These emission centres (ECs) are spatially confined regions that contain both charge and exciton trap states[23,24] and are single photon emitters[22]. By targeting a single exciton system with TR-STML, we show how this powerful technique reveals the details of exciton formation and decay. By controlling the charge injection at an EC, we systematically alter its luminescence response and derive parameters that lead to a charge dynamics model.

Fig. 1 shows a TR-STML scheme which couples fast voltage pulses with an STML setup comprised of a low-temperature (4 K) STM with an optical path to the tunnel junction[44]. This STML setup can monitor spatially-resolved light emission from the STM tip apex region, which is far below the diffraction limit, because it combines spatially localized current injection and tip-enhanced coupling between the near and far electromagnetic field. To perform time-resolved experiments, an arbitrary wave generator (AWG) delivers to the tunnel junction a train of 100 ns long rectangular voltage pulses corrected for the STM wiring transfer function[26] (cf. Methods). The pulses induce a time-dependent electroluminescent response from ECs in $C_{60}$ multilayers, which is recorded by a time-resolved single photon detector. The experimental time resolution of 1 ns is limited by the wiring cable bandwidth[26]. The photon detector time resolution is below 50 ps. All measurements are performed on 6-12 monolayers of a $C_{60}$ thin film epitaxially grown on Au(111).

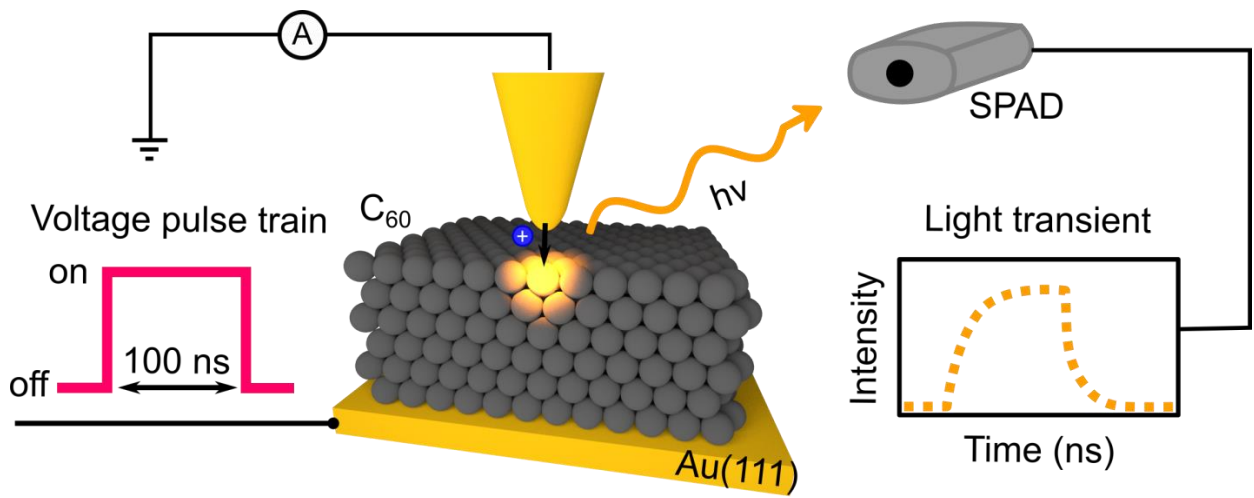

**Figure 1 | Time resolved STM-induced luminescence (TR-STML) setup.** 100 ns long voltage pulses with 2 MHz repetition rate are coupled into an STM junction consisting of a gold tip, a vacuum barrier, $C_{60}$, and a Au(111) substrate. Holes injected during the pulse induce electroluminescence at specific defects (ECs) in the $C_{60}$. The emitted photons trigger a single-photon avalanche photo diode (SPAD) with 50 ps time resolution. A histogram of counts at a defined delay (light transient) is obtained by synchronizing the photon detection time with the pulse arrival time.

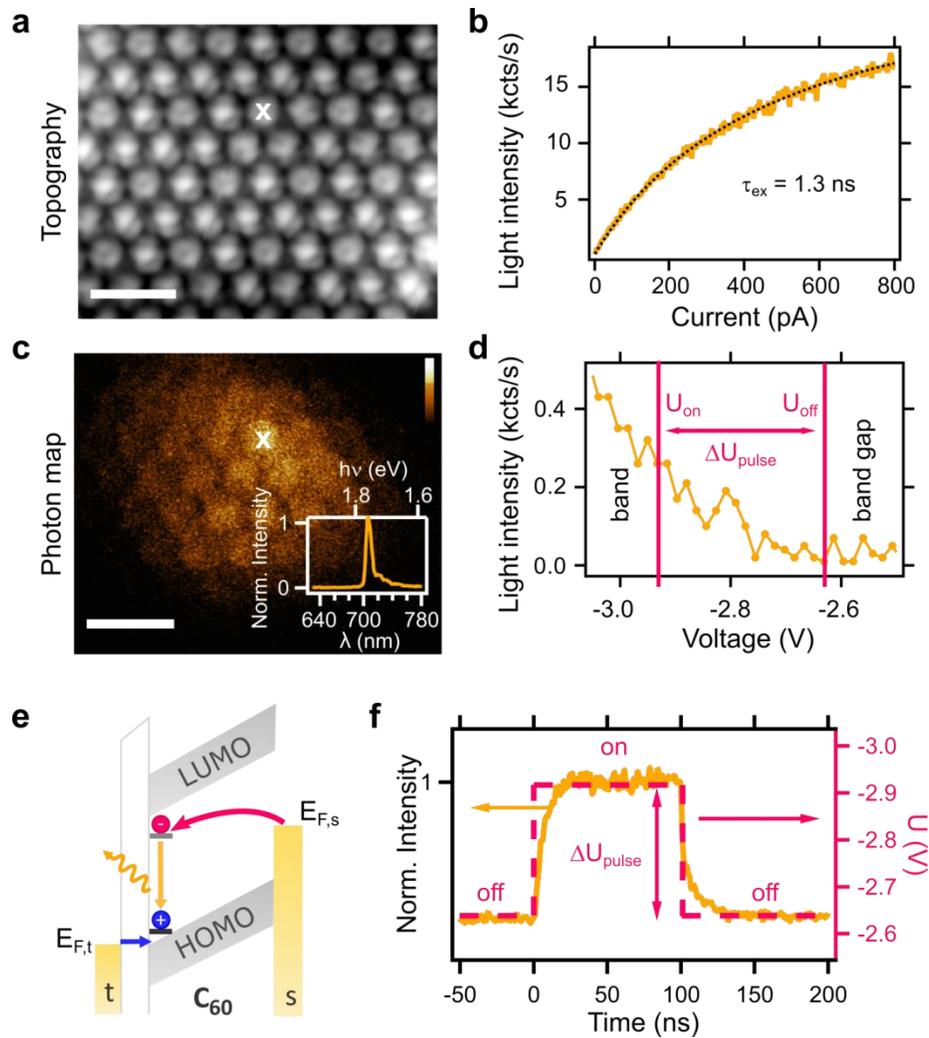

**Figure 2 | Characterization of an emission centre in a C$_{60}$ thin film. a**, Constant-current STM image of a C$_{60}$ layer, U$_{bias}$ = -3 V, I$_{tunnel}$ = 50 pA. The scale bar represents 2 nm. **b**, Luminescence versus current measurement at position X in figures **a** and **c**. The dashed curve is a fit using equation (1), which yields the exciton lifetime $\tau_{ex}$. **c**, Electroluminescence photon map recorded together with the image **a**. The vertical light intensity scale bar ranges from 0.0 to 5.3 kcounts s$^{-1}$; the horizontal length scale bar in the lower left corner represents 2 nm. The inset is a light spectrum recorded at position X at U$_{bias}$ = -3 V, I$_{tunnel}$ = 700 pA, integration time of 200 s. **d**, Luminescence vs. bias voltage measured at position X in **a** and **c**. Set point of the tunnel junction: -3 V, 5 pA. The magenta vertical lines indicate the bias voltage levels during (U$_{on}$ = -2.93 V) and between (U$_{off}$ = -2.63 V) pulses, $\Delta$U$_{pulse}$ = -0.3 V. The trace is not corrected for the dark counts of the detector (0.04 kcounts s$^{-1}$). **e**, Energy level diagram of the event sequence leading to light emission at the defect state: hole injection (blue arrow), electron injection (magenta arrow) and exciton recombination (orange arrow). **f**, applied voltage pulse (magenta dashed line) and measured time-resolved electroluminescence (orange line).

Fig. 2a shows the $C_{60}$ film surface imaged with STM at constant current condition. The simultaneously recorded photon map (Fig. 2c) reveals the spatial extent of the EC. Its luminescence spectrum (inset of Fig. 2c) obtained at the position marked by a white cross in Fig. 2c has a sharp feature with a vibrational progression that is due to a locally allowed $S_1$ to $S_0$ transition of $C_{60}$[23,43]. The particular EC defect shown in Fig. 2 is buried subsurface[23] and lacks any detectable electronic in-gap states (see Supplementary Fig. S2) which would otherwise be observable in scanning tunnelling spectra if a defect were located directly at the surface[24].

The basic electroluminescence mechanism of the EC is shown in Fig. 2e. Near -3 V, hole injection (blue arrow) into the band derived from the highest occupied molecular orbital (HOMO) of $C_{60}$ becomes possible. Any energetically shallow hole traps consisting of in-band-gap states will also be populated, resulting in hole trapping at spatially defined regions of the film. The positive charge, due to Coulomb attraction, lowers the energy of the lowest unoccupied molecular orbital (LUMO) and a corresponding electron trap state. Next, an electron can be injected from the substrate (magenta arrow) which partners up with the hole to form an exciton that is able to radiatively recombine (orange arrow). The electron injection through the energy barrier at the $C_{60}$-Au(111) interface occurs at moderate bias near -3 V only for films thicker than 4 monolayers. Otherwise, the potential drop within the $C_{60}$ film would be insufficient to shift both the LUMO-derived states and the trap to the Fermi energy of the substrate ($E_{F,s}$) and enable electron injection from the substrate even after hole trapping as shown in Fig. 2e[23].

**Exciton lifetime determination**

Exciton recombination dynamics under steady-state conditions are readily accessible without the use of sophisticated time-resolved detection schemes because of the conjugate relationship between the tunnelling current and the average time between injected charges. Fig. 2b shows the light intensity as a function of current up to 800 pA, expressed as the number of detected photons per second. At this current, the average time between consecutive tunnelling charges is $\tau_{tunnel}$ = 0.2 ns, which is comparable to, or smaller than the exciton lifetime of about 1 ns. The short time between the injected holes leads to non-radiative annihilation of the exciton[22]. The light intensity $P(\tau_{tunnel})$ behaves as[22]:

$$P(\tau_{tunnel}) = \eta \frac{\alpha}{\tau_{tunnel} + \beta \tau_{ex}} \qquad (1)$$

where $\alpha$, $\beta$, $\eta$ are parameters describing the charge trapping efficiency, exciton-charge annihilation, and photon detection, respectively. Fitting equation (1) to the data in Fig. 2b yields an exciton lifetime of $\tau_{ex}$= 1.3 ns, which is the same order of magnitude as the 0.75 ns measured directly in a previous time-domain photon-photon correlation experiment[22]. Obtaining the exciton lifetime from fitting light intensity versus current curves can thus serve for similar systems as a practical alternative to photon-photon correlation measurements on single excitons.

**Measurements of exciton formation dynamics**

Exciton formation dynamics can be probed by an explicit time-dependent experimental approach such as measuring the step response of the EC in the TR-STML scheme. Fig. 2d shows the light emission onset at the band edge near -2.7 V. By shifting the voltage in our experiment closer to -3 V, the Fermi energy of the tip $E_{F,t}$ shifts into the HOMO-derived band (Fig. 2e) so that a tunnelling current sets in and light is emitted. Therefore, it is sufficient to toggle between -2.93 V and -2.63 V (marked with two vertical lines in Fig 2d) to switch the luminescence on and off. This is achieved by adding a -0.3 V rectangular voltage pulse to a -2.63 V offset, shown in Fig. 2f in magenta. The orange trace represents the step response of the EC luminescence. The light intensity rises and reaches steady-state after 25 ns. After the pulse, $E_{F,t}$ returns into the bandgap and switches off the tunnel current. Crucially, the luminescence does not vanish instantaneously. After a rapid decrease, which will be discussed below, it decays exponentially with a time constant of 15 ns, whose origin *cannot* be the exciton lifetime because $\tau_{ex}$ is far shorter. The same analysis holds for the delayed rise of the transient. These observations show that TR-STML is sensitive to the slower processes involved in exciton formation. Recent single molecule STML studies report exciton formation due to plasmon energy absorption process[19,21,45]. This would suggest a coupling between tunnelling current and exciton formation on the femtosecond timescale of plasmon lifetime[26]. By contrast, the dynamics of $C_{60}$ ECs occurring on nanosecond timescales presents an irrefutable proof that the $C_{60}$ excitons originate from the fusion of two charges of opposite sign.

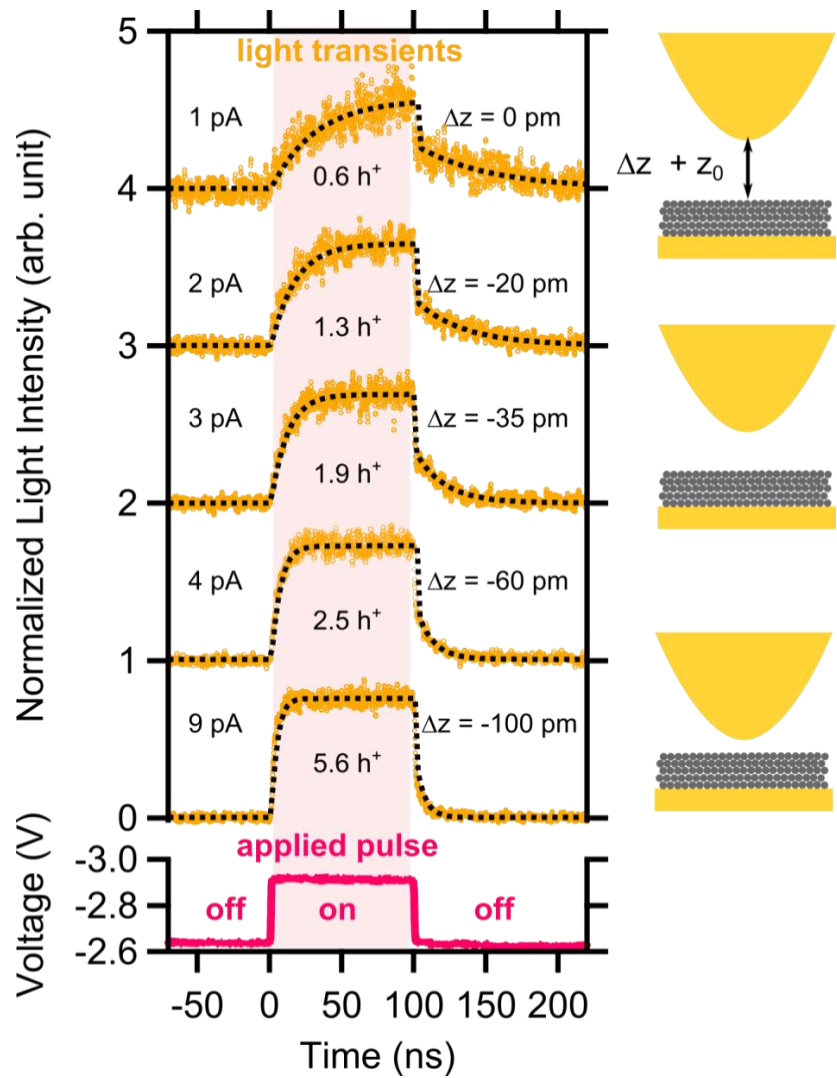

**Figure 3 | Electroluminescent response as a function of tip-sample distance.** Light transients (orange traces) as a function of the tip-sample distance, decreasing from top to bottom. The absolute distance is defined as $z = z_0 + \Delta z$, where $z_0$ is the tip-sample distance at U = -2.93 V and 1 pA tunnelling current (top trace). Traces are offset for clarity. Each trace is labelled from left to right by the tunnelling current I for corresponding z at $U_{on}$, the average number of holes ($h^+$) injected during one pulse (Q = I·$t_{pulse}$, $t_{pulse}$ = 100 ns), and the distance change $\Delta z$ with respect to the top trace. The full data set is presented in Supplementary Fig. S3. The dashed lines are fits to the model described in the main text. Bottom: Applied pulse shape (magenta trace) measured by plasmonic emission on a clean Au(111) surface and converted to the voltage in the junction. Note the difference between the pulse and the delayed luminescence responses. The pink shading indicates the time interval when the pulse is "on". The cartoon on the right illustrates the reduction of the tip-sample distance between the measurements (not to scale).

We have measured the electroluminescence transient of the EC at its brightest position (marked by a cross in Fig. 2d). These measurements are done at a defined tip-sample distance (z), which determines the tunnelling current and the electric field in the system. By varying z, we change both, thus systematically controlling the dynamics of the system as shown in Fig. 3. The topmost trace is recorded with the tip retracted to $z_0$, the furthest distance from the $C_{60}$ surface studied. The consecutive traces were measured with the tip repositioned by $\Delta z$. With decreasing z the charge injection rate grows exponentially, which increases the light intensity correspondingly. The tip-sample distance influence on the measured dynamics is apparent in Fig. 3. The closer the tip is to the surface, the steeper the step response becomes, which is evidence of faster dynamics. This behaviour is typical of all ECs investigated and does not show a simple dependence on the film thickness (see Supplementary Information). In Fig. 3, the average number of injected holes or extracted electron ranges from 0.6 charges per pulse (top trace) to 5.6 charges per pulse (bottom trace), indicating that our study reaches the single charge injection limit and the single exciton regime. The latter follows from the fact that the number of created excitons must be lower than the number of injected holes.

**Model of charge and exciton dynamics**

To analyse the origin of TR-STML transients and their tip-sample distance dependence we use a rate equation model describing the charge dynamics of exciton formation and recombination that includes their respective time constants ($\tau$). The transients can be divided into three time intervals: before, during, and after the pulse (Fig. 4), which shall be separately discussed. Before the pulse arrives to the tunnel junction (leftmost diagram), $E_{F,t}$ lies in the bandgap of $C_{60}$ ($U_{off}$).

Therefore, no charge is injected into the EC, no exciton created and no photon emitted. When the applied voltage changes to $U_{on}$, $E_{F,t}$ decreases below the HOMO-derived band edge. Therefore, a hole can be injected and trapped at the defect state (middle diagrams), occurring on average once every time interval $\tau_h$. A trapped hole shifts the electron trap level below $E_{F,s}$ due to Coulomb attraction, which enables an electron to be injected from the Au(111) substrate to the trap after a characteristic time $\tau_{el}^{on}$. Once this occurs, an exciton can be formed and can radiatively decay with its lifetime of $\tau_{ex} \approx 1.3$ ns. This sequence may occur more than once while the pulse is on. After 100 ns the pulse ends, but the luminescence persists for much longer than the exciton lifetime (right side of Fig. 4). The falloff at the end of the pulse consists of a rapid intensity drop on the order of the exciton lifetime, followed by a much slower decay. This decrease (see $\delta$ in Fig. 4 and Supplementary Information) reflects the quasi-instantaneous change of the time constants of the system in response to the falling edge of the voltage, which returns $E_{F,t}$ back to the bandgap. Since hole injection is forbidden after that transition, the continued light emission is direct proof of a trapped hole persisting beyond the pulse duration. This hole is responsible for shifting the electron trap energy level below $E_{F,s}$, so that electron injection is still possible with its characteristic time $\tau_{el}^{off}$. However, another process involving the trapped hole may occur: since after the pulse $E_{F,t}$ is energetically above the defect state, the hole can leave the trap and tunnel back to the tip, which we denote as "detrapping" with a characteristic time $\tau_d$. This process competes with electron injection (i. e. exciton creation) and reduces the electroluminescence after the pulse ends.

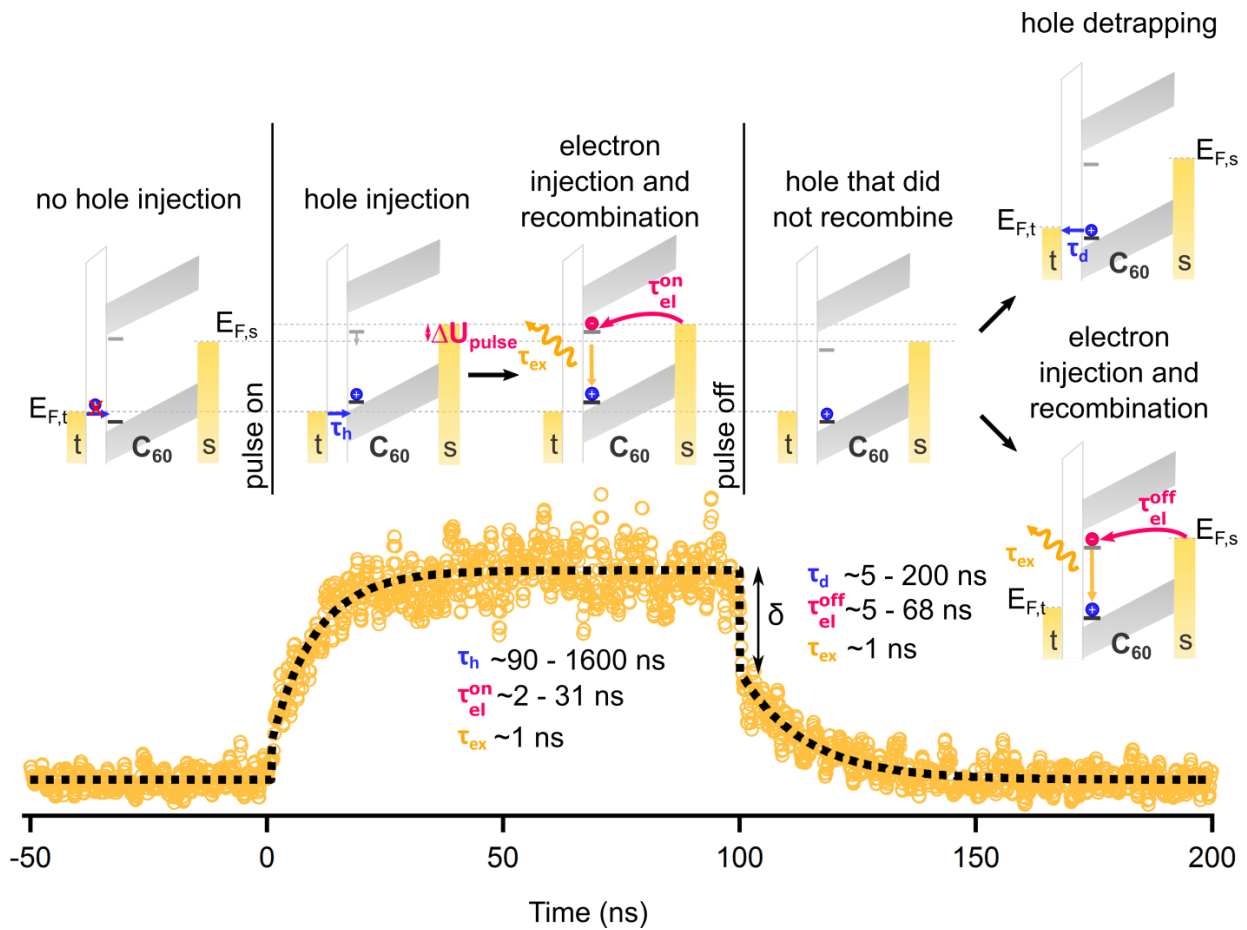

**Figure 4 | Model of exciton formation and recombination dynamics at the EC during a transient measurement.** The energy diagrams along the top of the figure show the system state during measurements at times before, during, and after the voltage pulse. Details of the model are described in the main text. Below the energy diagrams, a measured transient (orange circles) is given as an example; the black dotted line is a fit to the rate model. The typical ranges for the time constants are displayed next to the transient.

In principle, excitons could also be generated by injecting an electron first from the substrate, followed by a tunnelling hole. However, such an electron tunnelling to the LUMO-derived states can tunnel further to the tip and create an additional photon emission channel from inelastic tunnelling[23]. The absence of any corresponding broadband plasmonic emission in the inset of Fig. 2c is evidence to exclude this reverse injection mechanism as the origin of exciton formation at the EC studied. To strengthen this argument, we calculated the potential within the $C_{60}$ layer (cf. Methods) with and without a trapped hole, which is shown in Fig. 5. The hole lowers the LUMO-derived states by about 0.5 eV, demonstrating that the electron trap level is shifted below $E_{F,s}$ only in the presence of a hole, hence suppressing the reverse injection mechanism.

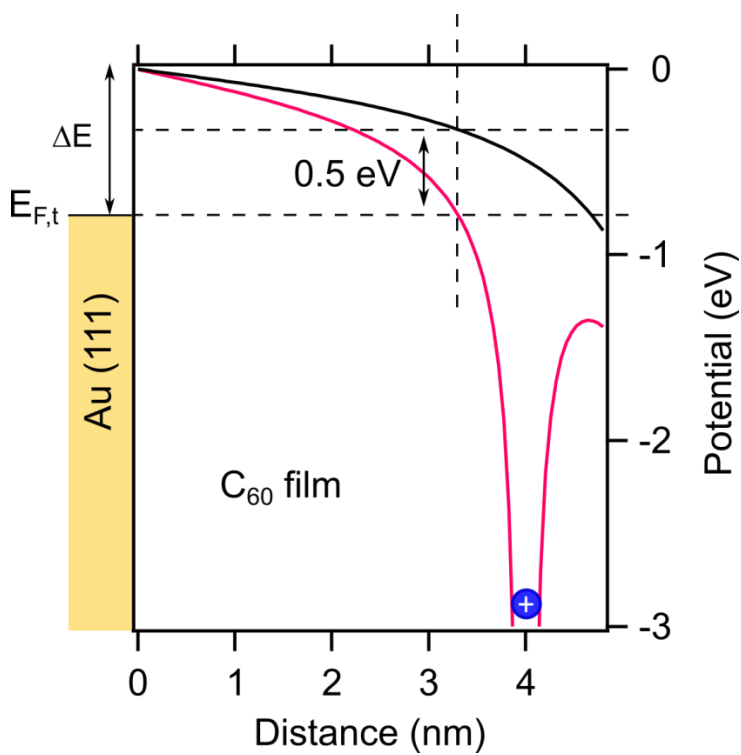

**Figure 5 | Electrostatic potential in the $C_{60}$ film.** The electric potential within the $C_{60}$ layer corresponding to the LUMO-derived band onset is calculated for bias U = -2.93 V with (magenta curve) and without (black curve) a trapped hole. The STM tip is assumed to be on the right side at a distance of 0.5 nm from the film surface. The presence of the hole lowers the potential barrier ($\Delta E$) within the $C_{60}$ film by around 0.5 eV and enables electron injection from the substrate into the film. Details of the calculation are discussed in Methods. The position of the LUMO-derived band with respect to $E_{F,t}$ is assumed to be 0.8 eV[46].

The dynamic behaviour as described can be analysed by a three-state kinetic model[20,22] whose analytic solutions are the time-dependent populations of the ground state $n_0(t)$ (empty defect), intermediate state $n_h(t)$ (trapped hole) and excited state $n_{ex}(t)$ (exciton). These population functions are a sum of weighted exponentials. Different equations apply during and after the pulse, making the respective exciton populations $n_{ex}^{on}(t)$ and $n_{ex}^{off}(t)$ both depend on the 5 rates $\frac{1}{\tau_{ex}}, \frac{1}{\tau_h}, \frac{1}{\tau_d}, \frac{1}{\tau_{el}^{on}}, \frac{1}{\tau_{el}^{off.}}$. The observable time-dependent light intensity $P(t)$ is always proportional to the exciton population $n_{ex}(t)$. A derivation for $P(t)$ can be found in the Supplementary Information. Two of the rates are known from steady-state results: the reciprocal exciton lifetime $\frac{1}{\tau_{ex}}$, and the hole injection rate obtained from the injection current (I) during the pulse $\frac{1}{\tau_h} = \frac{I}{\alpha e}$ assuming a hole trapping efficiency of $\alpha \approx 10\%$. Hence, there are 3 free parameters $\left( \frac{1}{\tau_d}, \frac{1}{\tau_{el}^{on}}, \frac{1}{\tau_{el}^{off.}} \right)$ that can be obtained from fitting the transients in Fig. 3. Also the transients can be parameterized with 3 independent parameters, namely, the rising and falling edge time constants, and the immediate fall-off $\delta$ (see Fig. 4). These fitting parameters are related to the 3 free parameters of the model. Using $P(t)$ we can excellently fit the transient electroluminescence trains (Fig. 3) and extract a set of time constants. Their typical ranges are listed in Fig. 4. The slowest process is hole injection by the STM tip ($\tau_h$). As it determines the maximum exciton population, the light intensity is the most sensitive to this parameter. The electron injection ($\tau_{el}^{on}$) is a faster process, as Coulomb attraction between trapped hole and charge carriers from the substrate leads to efficient exciton generation. The detrapping process ($\tau_d$) is found to induce only a minor effect on the traces. The corresponding rate $\frac{1}{\tau_d}$ as a function of the tip-sample distance has an exponential dependence (see Supplementary Fig. S4) that is consistent with our interpretation of the trapped hole tunnelling back to the tip. At currents around 10 pA, the charge dynamics of the system becomes fast and the transient response becomes increasingly sharp, with rounded edges due to the finite exciton lifetime and electron injection time being on the order of 1 ns. For currents beyond the measured range the exciton lifetime alone would dominate the transient behaviour at the pulse boundaries. We note that the model can also be extended to the case where the offset voltage is located in the HOMO-derived band, so that the pulse only modulates the hole injection rate (see Supplementary Fig. S5).

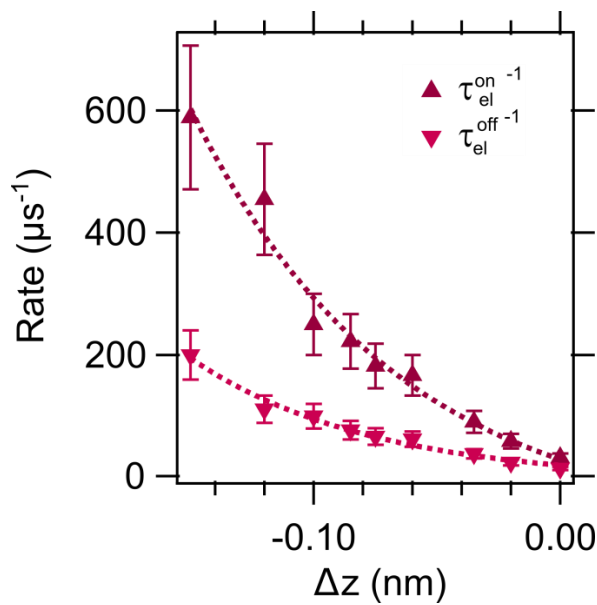

**Figure 6 | Electron injection dynamics analysis.** The rates (inverses of time constants) describing electron dynamics are plotted as a function of the tip-sample distance change ($\Delta z$). The dashed exponential curves are guides to the eye. The error in the time constants obtained from the fits to the model is ±10%.

Fig. 6 presents as a function of $\Delta z$ the electron injection rates $\left(\frac{1}{\tau_{el}^{on}}, \frac{1}{\tau_{el}^{off.}}\right)$ obtained by fitting the transients in Fig. 3 to the model. The electric field in $C_{60}$ is approximately linear with the tip-sample distance. Hence, the observed non-linear dependence of electron injection on $\Delta z$ in Fig. 6 is a confirmation of the presence of an energy barrier at the interface that results in a strong non-ohmic character of the Au(111)/$C_{60}$ junction. The charge injection through this barrier is affected by subsurface defects (e.g. roughness of the Au(111) surface) and hidden interface dynamics (e.g. interface dipoles and charge transfer)[47,48]. Our results demonstrate that TR-STML is sensitive to these hidden effects at the single-electron limit and is a valuable testbed for the development of a more detailed understanding of the fundamentals of charge injection at the quantum level in technologically relevant interfaces.

In conclusion, TR-STML is a general method that temporally and spatially resolves individual quantum processes, as shown for single exciton formation and recombination. In this work, we control the response of a model OLED system by locally altering the charge injection dynamics at the organic-metal interface. We have used this new approach to investigate single ECs in $C_{60}$ and have determined all relevant time constants, including the dynamics of charge injection and exciton lifetime from these experiments. By overcoming limitations of methods that rely on ensemble measurements and high excitation densities, TR-STML opens up new possibilities for studying how fundamental processes drive the conversion between charges and photons. TR-STML is naturally suited to real-space mapping of charge and exciton dynamics at the ns-nm scale in systems such as single fluorophores, molecular layers, and 2D materials, as well as solid-state photon emitters. Applying this technique to these systems will help pinpoint and optimize the many key parameters in optoelectronic devices.

**Acknowledgements.**

We thank S. Loth for fruitful discussions and R. Froidevaux for help with the experimental set-up. A.R. thanks T. Michnowicz for valuable comments on the manuscript. P.M. acknowledges the support of the A. v. Humboldt Foundation.

**Author contribution.**

A.R., P.M., C.G. and C.C.L. performed the experiments, analysed the data and discussed the model. O.G. and A.R. performed the electrostatic calculations. M.E., K.Ku., and K.Ke. conceived of and supervised the project. A.R. wrote the manuscript with input from all authors.

**Competing financial interests.**

The authors declare no competing financial interests.

**Methods.**

**Sample fabrication.** The Au(111) single crystal substrate is cleaned under UHV conditions with repeated cycles of Ar$^+$ ion sputtering followed by annealing at 850 K. C$_{60}$ molecules are thermally evaporated for 1 hour from a crucible at 840 K onto the substrate kept at room temperature. The sample is then transferred *in situ* into the liquid-He cooled UHV STM. STM tips are prepared by electrochemical etching of an Au wire of 99.995% purity. Before measuring on C$_{60}$, the tips are prepared on a pristine Au(111) surface with voltage pulses and controlled surface indentations.

**Scanning tunnelling microscopy.** All experiments are performed with a home-built, UHV ($<10^{-10}$ mbar), low temperature (4 K) STM. The light emitted from the tunnel junction is collected by three lenses surrounding the STM tip and guided to three detectors located outside the vacuum chamber[44]. One optical path leads to a spectrograph (Acton SP 300i) coupled to an intensified CCD camera (PI-MAX), while the other two lead to single-photon avalanche photo diode (SPAD) modules (MPD-PDM-R) with 50 ps time resolution. In this work, just one of the SPADs is used. The STM tip is grounded and the bias voltage plus the nanosecond pulses are applied to the sample.

**Time-resolved STML.** A continuous train of 100 ns long voltage pulses (2 MHz repetition rate) is produced by an arbitrary wave generator (AWG, Agilent M8190A) and fed to the STM head through semi-rigid and coaxial high frequency wires. In order to correct for the imperfections of the wiring the voltage pulses are shaped[26] such that rectangular pulses with sharp edges arrive at the junction (see Supplementary Fig. S6). With this compensation the 10%-90% voltage transition times of both the rising and falling edges are 1 ns. The pulses have a height of −300 mV and are added to the -2.63 V DC offset voltage with a bias tee (Picosecond Pulse Labs, 5550B). The time-dependent electroluminescence intensity is recorded with a time-correlated single photon counting PC card (Becker & Hickl, SPC-130) which measures the time interval between a trigger pulse from the AWG and the next photon detected by the SPAD. The number of occurrences of each time interval is rendered as a histogram which provides the electroluminescence transient in the limit of low intensity which is fulfilled for all data shown. During a measurement (20 minutes integration time) more than one billion voltage pulses arrive to the tunnel junction, but only a small fraction of pulses results in a photon detection event because of the low excitation and collection efficiency of $10^{-5}$ photons per charge.

**Electrostatic calculation.** To calculate the potential in the $C_{60}$ film, we use the image charge method and treat the layer as a homogeneous dielectric medium. The system consists of a sphere (tip) with a radius of r = 3 nm (see Supplementary Information for more details), a vacuum gap of d = 0.5 nm, 6 layers of $C_{60}$ with a thickness of a = 4.8 nm and a relative permittivity $\varepsilon_r$ = 4.4, and a metal electrode. The image charges are iteratively added at the tip, vacuum, $C_{60}$ layer and metal until boundary conditions at all interfaces converge.

**Supplementary Information for**

# Single charge and exciton dynamics probed by molecular-scale-induced electroluminescence

A. Rosławska[1]*, P. Merino[1,+], C. Große[1,#], C. C. Leon[1],  O. Gunnarsson[1], M. Etzkorn[1,‡], K. Kuhnke[1], K. Kern[1,2]

[1] Max Planck Institute for Solid State Research, Heisenbergstraße 1, 70569, Stuttgart, Germany.

[2] Institut de Physique, École Polytechnique Fédérale de Lausanne, 1015 Lausanne, Switzerland.

+ present address: Instituto de Ciencia de Materiales de Madrid, CSIC, c/Sor Juana Inés de la Cruz 3, E28049, Madrid, Spain.

# present address: NanoPhotonics Centre, Cavendish Laboratory, University of Cambridge, Cambridge CB3 0HE, UK.

‡ present address: Institut für Angewandte Physik, TU Braunschweig, Mendelssohnstraße 2, 38106 Braunschweig, Germany

* corresponding author: a.roslawska@fkf.mpg.de

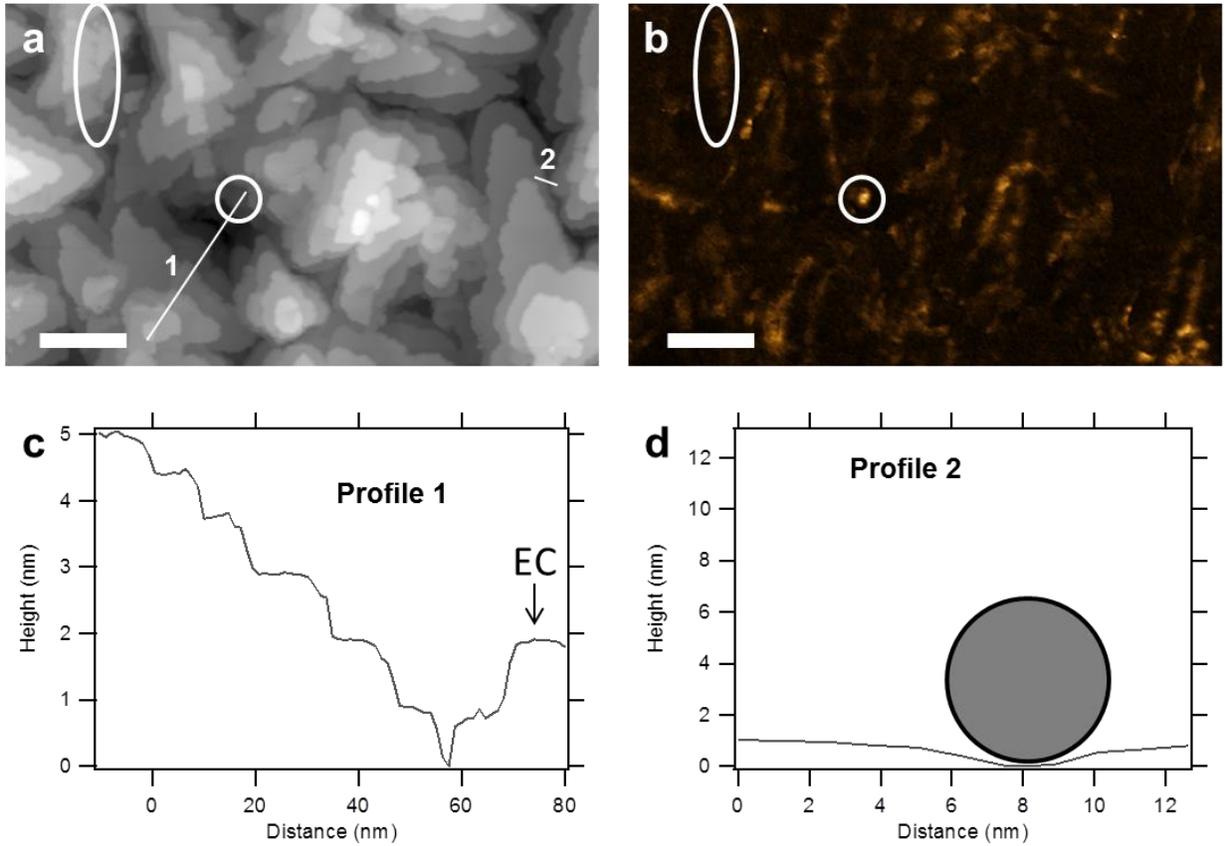

**Supplementary Figure S1. Large scale image analysis of the C$_{60}$ film. a**, Topography. **b**, Simultaneously recorded photon map. The white circles mark the EC shown in detail in Fig. 2. The white ovals indicate a domain boundary with ECs. Scale bar 20 nm. **c, d**, Line profiles at regions marked by 1 and 2 in **a**, respectively. The circle in **d** has a radius r = 3 nm and represents the largest possible tip curvature, which still allows recording the measured profile.

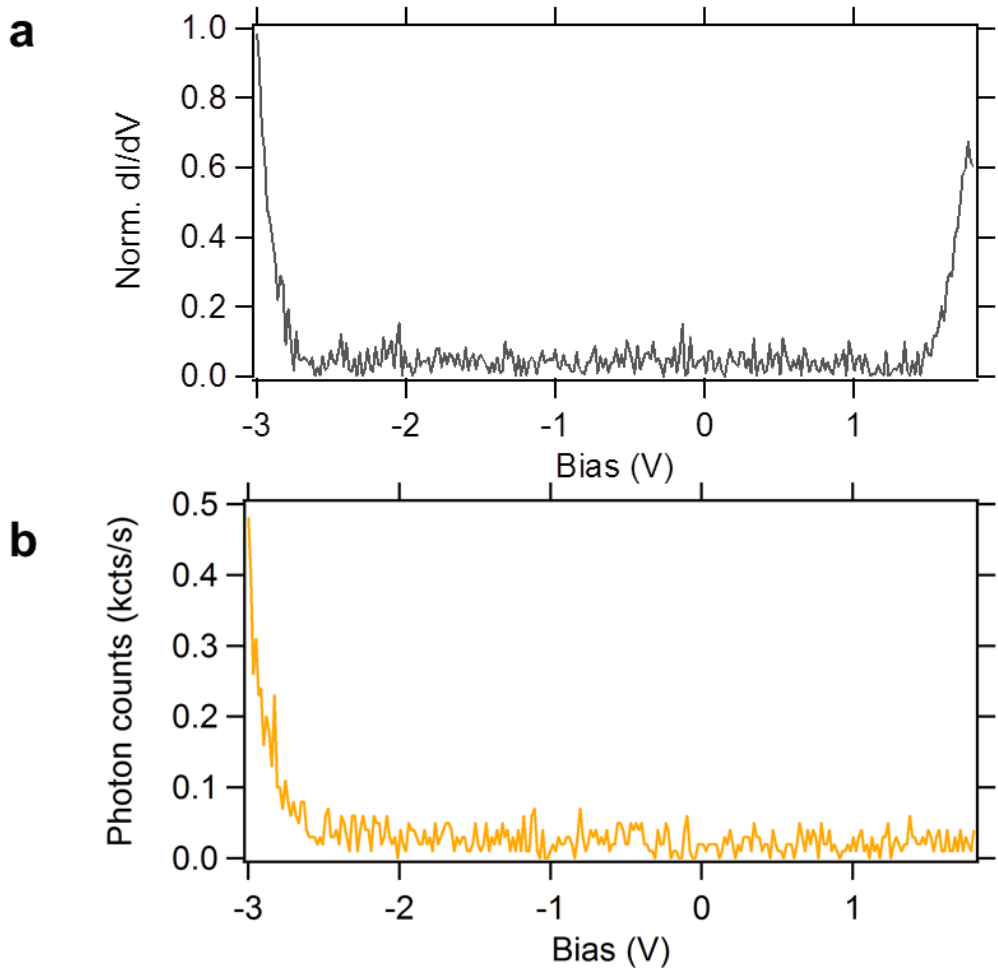

**Supplementary Figure S2. Scanning tunnelling spectroscopy recorded at the investigated emission centre. a**, dI/dV spectrum, in which shallow trap states are not observed. **b**, Simultaneously recorded light intensity vs. applied bias showing that light is emitted only at negative polarity. The count rate does not drop to zero due to the dark count rate of the detector (0.04 kcts/s). Set point of the tunnel junction: -3 V, 5 pA.

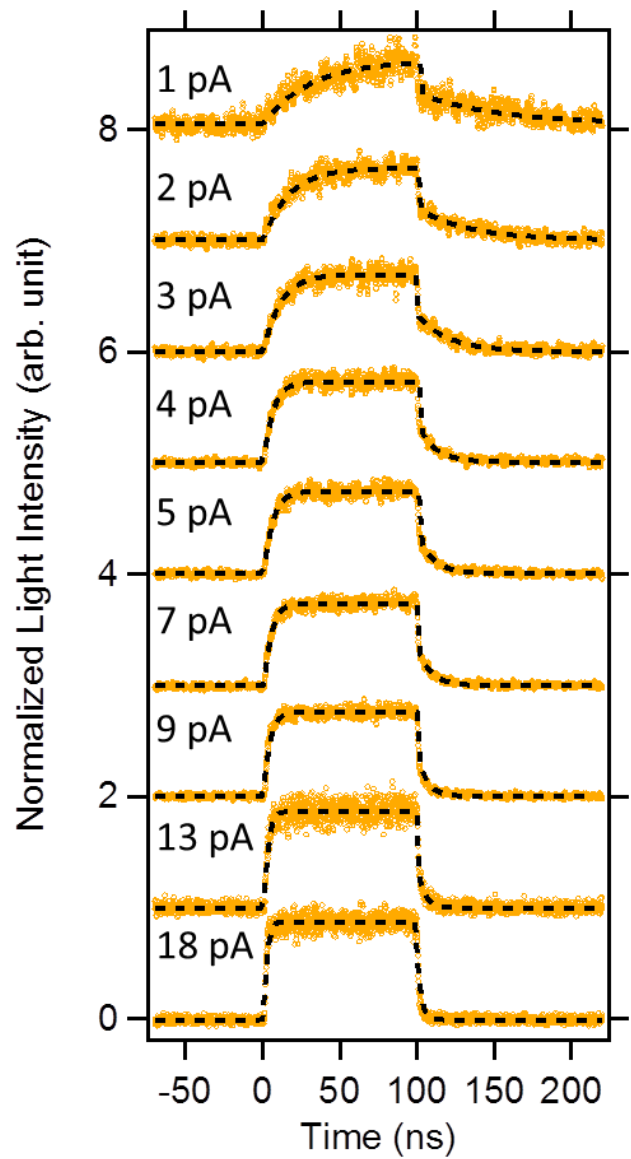

**Supplementary Figure S3. Full dataset measured on the emission centre.** $U_{on}$ = -2.93 V, $U_{off}$ = -2.63 V. Current set points are indicated next to the traces. Dashed lines are fits to the data based on the rate equation model. Traces are offset for clarity by multiples of one half of a tick mark spacing. Fig. 3 of the main manuscript presents selected traces.

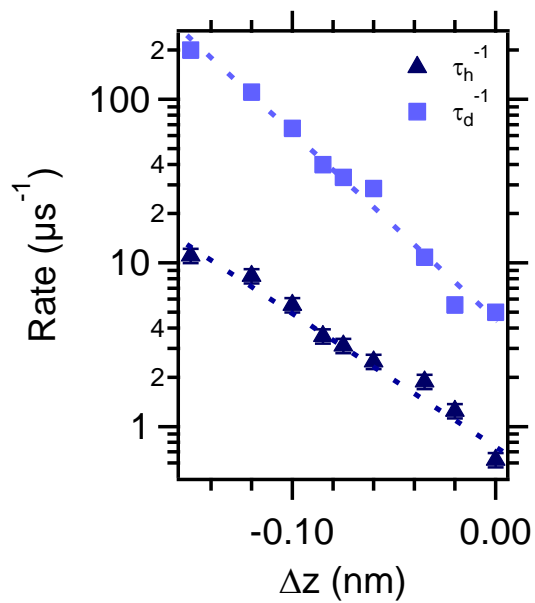

**Supplementary Figure S4. Hole-related time constants plotted on a logarithmic scale.** Squares represent the detrapping rate obtained from fitting the measured transients as described in the main text. Triangles refer to the hole injection rate and involve the hole trapping efficiency. Dashed lines are guides to the eye (exponentials).

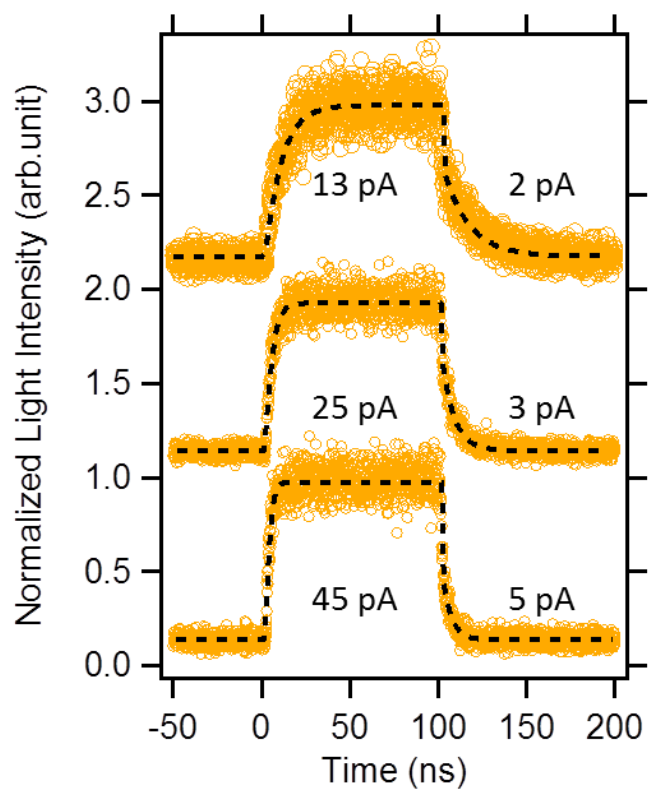

**Supplementary Figure S5. Measurements within the HOMO-derived band.** $U_{on}$ = -3.13 V, $U_{off}$ = -2.83 V. Currents during and after the pulse are indicated next to the traces. Traces are offset for clarity by multiples of two tick mark spacings. Black traces are fits to the model extended to measurements in which the voltage is always toggled within the HOMO-derived band, so that the charges are always being injected. To model the transients in this case, we include hole injection for both phases (during the pulse and in-between pulses). The model fits the transients perfectly assuming known hole injection (current) and allowing the electron injection rates to be derived from it.

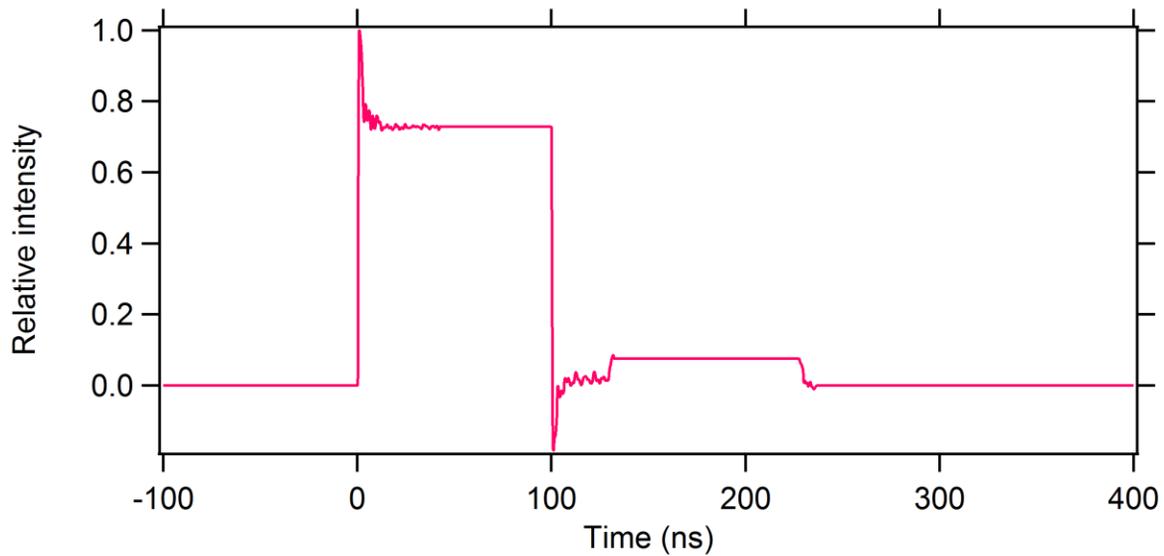

**Supplementary Figure S6. The pulse sent to the tunnel junction from the arbitrary wave generator.** The shape of the pulse accounts for the imperfections of the transfer function of the wiring. The pulse is designed so that it arrives rectangular at the tunnel junction (see Fig. 3, bottom trace). The procedure to shape the pulse has been described in an earlier publication[1].

**Thickness estimation**

The layer thickness can be estimated from a line profile across a large scale scan (Supplementary Fig. S1c). Emission centres are found on films with a thickness of at least 5 ML. Since the lowest visible terrace is dark (no electroluminescence) it has to be below 5 ML. The EC investigated in the Fig. 2 of the main text is located in a relatively low layer. Few emission centres can be found in the lower layer, therefore the EC is located at the 6 ML level.

**Charge dynamics as a function of the local structure**

The dynamics of the charge injection strongly depends on the number of layers, the number of nearby defects, and the roughness of the supporting metal surface. All of these parameters affect the electronic barrier for electron injection. For example, emission centres exhibiting fast dynamics at large tip-sample distances indicate a low injection barrier at the $Au(111) - C_{60}$ interface. Such a low barrier can arise due to other defects present below the emission centre, local topographic variations of the supporting metal substrate, a low number of layers, and local electronic effects at the interface.

**Tip radius**

The tip radius for electrostatic calculations has been estimated to be r = 3 nm. It is consistent with the ability of the tip to resolve narrow valleys and sharp steps in the z dimension (Supplementary Fig. S1d). The observed topographic structures on $C_{60}$ films can only be obtained by very sharp tips. With a blunt tip, the narrow valleys would exhibit a reduced depth.

**Details of the kinetic model**

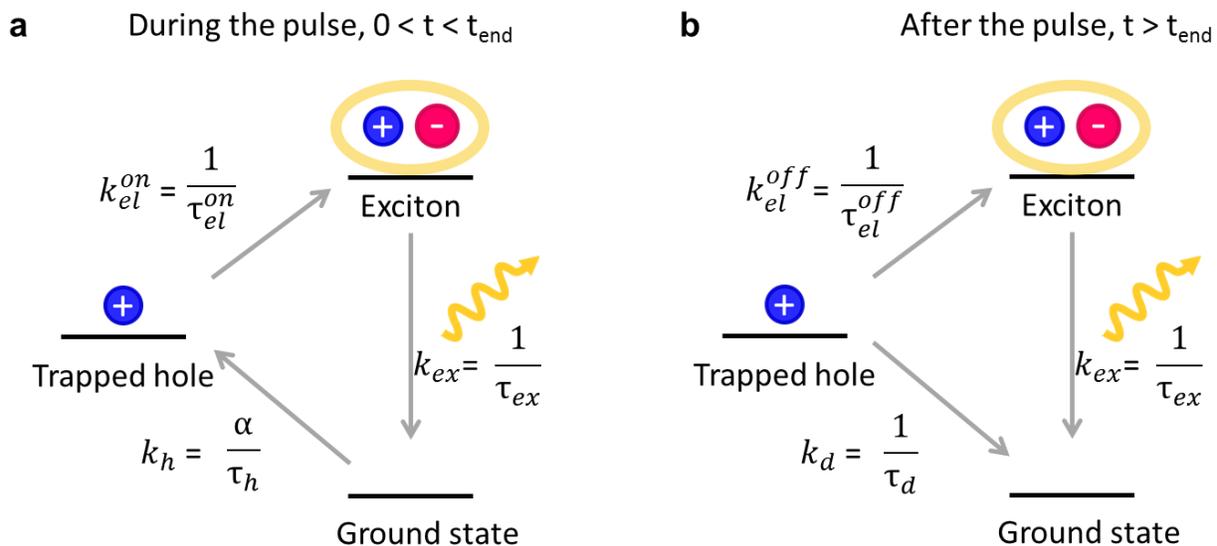

**a**        During the pulse, $0 < t < t_{end}$        **b**        After the pulse, $t > t_{end}$

**Supplementary Figure S7. Schematics of the kinetic model. a** and **b** show the processes with their corresponding rates (k) and time constants ($\tau$) during and after the pulse, respectively.

The model presented in the main text can be described in terms of a kinetic model using rate equations[2]. The rates (inverses of time constants) are presented schematically in Supplementary Fig. S7. Such a model is represented by the following coupled equations:

| During the pulse | After the pulse |
|---|---|
| $$\frac{dn_0^{on}}{dt} = n_{ex}^{on} k_{ex} - n_0^{on} k_h$$ $$\frac{dn_h^{on}}{dt} = n_0^{on} k_h - n_h^{on} k_{el}^{on}$$ $$\frac{dn_{ex}^{on}}{dt} = n_h^{on} k_{el}^{on} - n_{ex}^{on} k_{ex}$$ | $$\frac{dn_0^{off}}{dt} = n_{ex}^{off} k_{ex} + n_h^{off} k_d$$ $$\frac{dn_h^{off}}{dt} = -n_h^{off} k_{el}^{off} - n_h^{off} k_d$$ $$\frac{dn_{ex}^{off}}{dt} = n_h^{off} k_{el}^{off} - n_{ex}^{off} k_{ex}$$ |

With the following boundary conditions:

| During the pulse | After the pulse |
|---|---|
| $n_0^{on}(t = 0) = 1$ $n_h^{on}(t = 0) = 0$ $n_{ex}^{on}(t = 0) = 0$ | $n_0^{off}(t_{end}) = n_0^{on}(t_{end})$ $n_h^{off}(t_{end}) = n_h^{on}(t_{end})$ $n_{ex}^{off}(t_{end}) = n_{ex}^{on}(t_{end})$ |

They can be solved analytically[3] using an equating coefficients method for both sets of equations - during and after the pulse. The solutions to the time-dependent state populations have the form of weighted exponentials:

$$n(t) = Ae^{\lambda_1 t} + Be^{\lambda_2 t} + Ce^{\lambda_3 t}$$

The time-dependent detected light intensity P(t), defined for the three time intervals of the transient, depend on the exciton population, the exciton decay rate ($k_{ex}$) and the photon collection efficiency η. It is given by:

$$P(t) = \begin{cases} 0 \text{ for t<0} \\ \eta k_{ex}\, n_{ex}^{on}(t) \text{ for } 0<t<t_{end} \\ \eta k_{ex}\, n_{ex}^{off}(t) \text{ for } t>t_{end} \end{cases}$$

Where $n_{ex}^{on}$ and $n_{ex}^{off}$ are:

$$n_{ex}^{on}(t) = \eta\,\frac{W}{R}\,\left(1 + \frac{S-Q}{2\,Q}\cdot e^{-\frac{S+Q}{2}t} - \frac{S+Q}{2\,Q}\cdot e^{-\frac{S-Q}{2}t}\right)$$

$$n_{ex}^{off}(t) = \frac{n_h^{on}(t_{end})\cdot k_{el}^{off}}{k_{ex} - k_{el}^{off} - k_d}\, e^{-\left(k_{el}^{off}+k_d\right)t} + \left(n_{ex}^{on}(t_{end}) - \frac{n_h^{on}(t_{end})\cdot k_{el}^{off}}{k_{ex} - k_{el}^{off} - k_d}\right)e^{-k_{ex}t}$$

With:

$$\Omega = k_{ex}\cdot k_{el}^{on}\cdot k_h$$

$$S = k_{ex} + k_{el}^{on} + k_h$$

$$R = k_{ex}\cdot k_{el}^{on} + k_{ex}\cdot k_h + k_{el}^{on}\cdot k_h$$

$$Q = \sqrt{S^2 - 4R}$$

$$G = \frac{k_{ex}\cdot k_h}{R}$$

$$H = -\frac{1}{2RQ^2} \cdot k_h \left( -2\Omega + k_{ex}^3 - k_h^2 \cdot k_{el}^{on} - k_h k_{el}^{on\,2} - k_{ex}^2 \cdot Q^2 + k_{el}^{on} \cdot k_h \cdot Q \right.$$

$$\left. + R(-2k_{ex} + k_{el}^{on} + k_h + Q) \right)$$

$$J = \frac{1}{2RQ^2} \left( 2R^2 + \Omega(k_{ex} - k_{el}^{on} + k_h + Q) + k_{el}^{on}\left( k_{ex}^3 + k_{ex}^2 \cdot Q + k_h^2 \cdot (k_h + Q) \right) \right.$$

$$\left. - R\left( k_{ex}^2 + k_h(k_h + 2k_{el}^{on} - Q) + k_{ex}(2k_{el}^{on} + Q) \right) \right)$$

$$n_h^{on}(t_{end}) = G + H e^{\frac{-(S+Q)}{2t_{end}}} + J e^{\frac{-(S-Q)}{2t_{end}}}$$

$$n_{ex}^{on}(t_{end}) = \frac{k_h \cdot k_{el}^{on}}{R} \left( 1 + \frac{S-Q}{2Q} e^{\frac{-(S+Q)}{2t_{end}}} - \frac{S+Q}{2Q} e^{\frac{-(S-Q)}{2t_{end}}} \right)$$

The hole trapping efficiency α value must lie between 1 and $10^{-6}$. The lower bound is based on the experimentally measured light intensity above background being nonzero. Since not every injected hole is trapped, α has to be below 1. Here we assume α = 0.1. Since the resulting time constant ($\frac{\tau_{ex}}{\alpha}$) becomes long (being on the order of one μs) it has a negligible effect on the pulse shape and contributes only to the light intensity. α therefore can also take smaller values (longer time constant). Photon detection efficiency is estimated to be $\eta \approx 10^{-5}$.

The intensity drop described by the parameter δ (Fig. 4) is determined by the $\frac{\tau_{el}^{off}}{\tau_{el}^{on}}$ ratio. It appears quasi-instantaneous (< 2 ns) since the time constant for the electron injection change $\left( \tau_{el}^{on} \rightarrow \tau_{el}^{off} \right)$ occurs within 1 ns, which is the time-resolution limit of the measurement (edge of the pulse) and the decay time of an exciton is also on the order of 1 ns.

# Supplementary References